\documentclass[11pt, onecolumn]{article}
\usepackage[top=1in, bottom=1in, left=1in, right=1in]{geometry}

\usepackage{graphicx}
\usepackage{subfigure}
\usepackage{cite}
\usepackage{amsmath, amssymb}

\begin{document}
        \title{Modeling a Sensor to Improve its Efficacy}
   \author{Nabin K. Malakar\\
                W. B. Hanson Center for Space Sciences\\
                University of Texas at Dallas\\
                Richardson TX \\
                nabin.malakar@utdallas.edu\\
            \and
                Daniil Gladkov\\
                Department of Physics\\
                University at Albany (SUNY)\\
                Albany NY \\
                dgladkov@albany.edu\\
            \and
                Kevin H. Knuth\\
                Departments of Physics and Informatics\\
                University at Albany (SUNY)\\
                Albany NY \\
                kknuth@albany.edu}


    \maketitle
    \begin{abstract}
Robots rely on sensors to provide them with information about their surroundings.  However,
high-quality sensors can be extremely expensive and cost-prohibitive.  Thus many robotic systems must
make due with lower-quality sensors.  Here we demonstrate via a case study how modeling a sensor can improve its
efficacy when employed within a Bayesian inferential framework.  As a
test bed we employ a robotic arm that is designed to autonomously take
its own measurements using an inexpensive LEGO light sensor to
estimate the position and radius of a white circle on a black field.
The light sensor integrates the light arriving from a spatially
distributed region within its field of view weighted by its Spatial
Sensitivity Function (SSF).  We demonstrate that by incorporating an accurate model of the light sensor SSF
into the likelihood function of a Bayesian inference engine, an autonomous system can make improved
inferences about its surroundings.  The method presented here is data-based, fairly
general, and made with plug-and play in mind so that it could be
implemented in similar problems.
    \end{abstract}

\section{Introduction}
Robots rely on sensors to provide them with information about their surroundings.  However, high-quality sensors can be cost-prohibitive and often one must make due with lower quality sensors.  In this paper we present a case study which demonstrates how employing an accurate model of a sensor within a Bayesian inferential framework can improve the quality of inferences made from the data produced by that sensor.  In fact, the quality of the sensor can be quite poor, but if it is known precisely how it is poor, this information can be used to improve the results of inferences made from the sensor data.

To accomplish this we rely on a Bayesian inferential framework where a machine learning system considers a set of hypotheses about its surroundings and identifies more probable hypotheses given incoming sensor data.  Such inferences rely on a likelihood function, which quantifies the probability that a hypothesized situation could have given rise to the data.  The likelihood is often considered to represent the noise model, and this inherently includes a model of how the sensor is expected to behave when presented with a given stimulus.  By incorporating an accurate model of the sensor, the inferences made by the system are improved.

As a test bed we employ an autonomous robotic arm developed in the Knuth Cyberphysics Laboratory at the University at Albany (SUNY).  The robot is designed to perform studies in autonomous experimental design \cite{knuth_arm_2007,knuth_center2010}.  In particular it performs autonomous experiments where it uses an inexpensive LEGO light sensor to estimate the position and radius of a white circle on a black field.  The light sensor integrates the light arriving from a spatially distributed region within its field of view weighted by its Spatial Sensitivity Function (SSF).  We consider two models of the light sensor.  The na\"{i}ve model predicts that the light sensor will return one value on average if it is centered over a black region, and another higher value on average if it is centered on a white region.  The more accurate model incorporates information about the SSF of the light sensor to predict what values the sensor would return given a hypothesized surface albedo field.  We demonstrate that by incorporating a more accurate model of the light sensor into the likelihood function of a Bayesian inference engine, a robot can make improved inferences about its surroundings.  The efficacy of the sensor model is quantified by the average number of measurements the robot needs to make to estimate the circle parameters within a given precision.

There are two aspects to this work.  First is the characterization of the light sensor, and second is the incorporation of the light sensor model into the likelihood function of the robot's machine learning system in a demonstration of improved efficacy.  In the Materials and Methods Section we describe the robot, its light sensor, the experiment that it is designed to perform, and the machine learning system employed.  We then discuss the methods used to collect data from the light sensor, the models used to describe the light sensor, their incorporation into the machine learning system, the methods used to estimate the model parameters and select the model order.  The Results and Discussion Section describe the resulting SSF model and the results of the experiments comparing the na\"{i}ve light sensor model to the more accurate SSF model.  In the Conclusion we summarize our results which demonstrate how by incorporating a more accurate model of sensor, one can improve its efficacy.

\section{Materials and Methods}
In this section we begin by discussing various aspects of the robotic test bed followed by a discussion of the techniques used to characterize the light sensor.

\subsection{Robotic Arm Test Bed}
\begin{figure}[tb]
	\centering
		\includegraphics[width = 0.40\columnwidth]{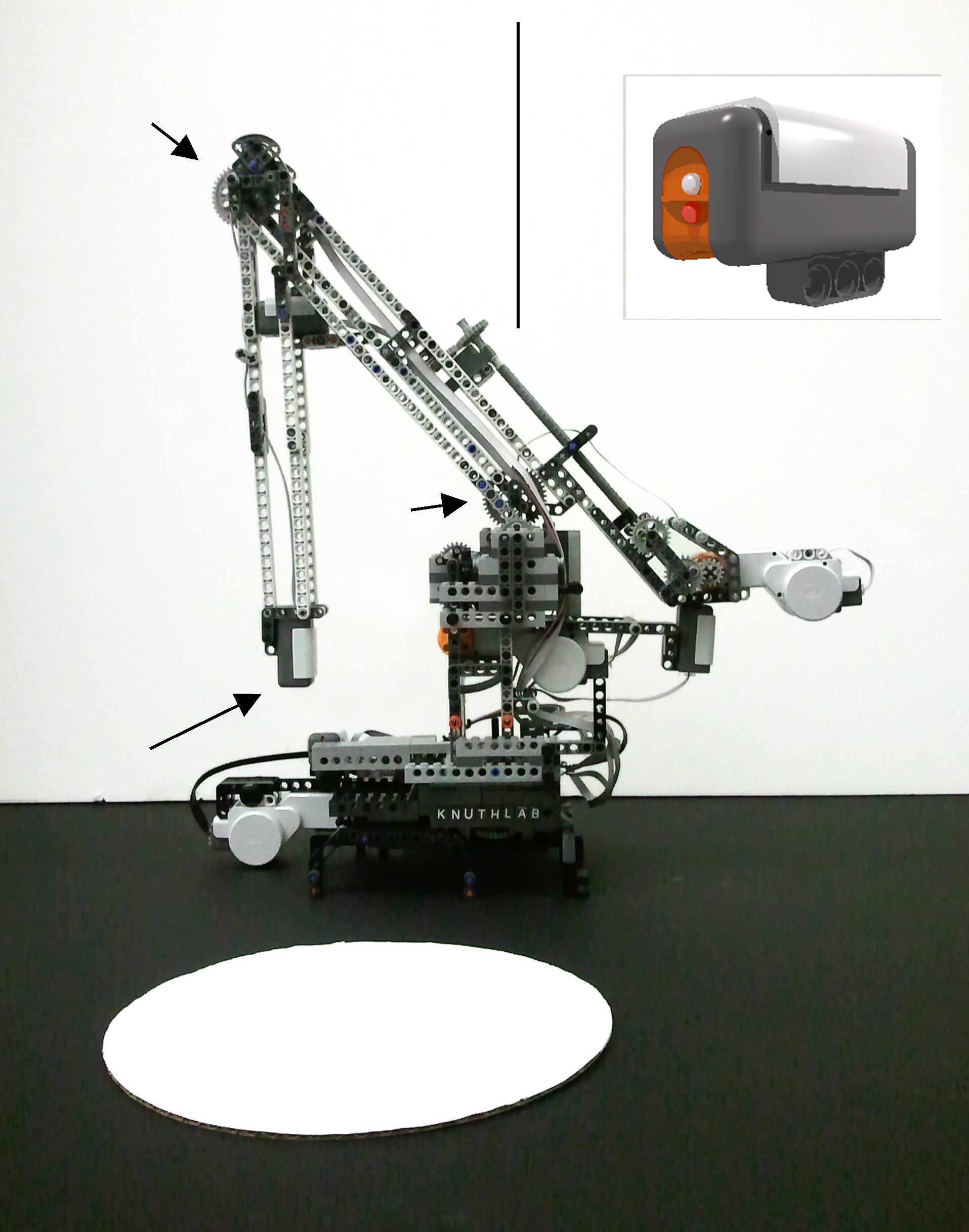}
	\caption[The Knuth Cyberphysics Robotic Arm]{A photograph showing the robotic arm along with the circle it is programmed to characterize. The robotic arm is constructed using the LEGO NXT Mindstorms System.  It employs one motor to allow it to rotate about a vertical axis indicated by the black line in top center of image, and two motors to extend and lower the arm about the joints located at the positions indicated by the short arrows.  The LEGO light sensor, also shown in the inset, is attached to the end of the arm as indicated by the long arrow.}
	\label{fig:arm-sensor}
\end{figure}

The robotic arm is designed to perform studies in autonomous experimental design \cite{knuth_arm_2007,knuth_center2010}.   The robot itself is constructed using the LEGO NXT Mindstorms system (Figure \ref{fig:arm-sensor}) \footnote{The LEGO Mindstorm system was utilized in part to demonstrate that high-quality autonomous systems can be achieved when using lower-quality equipment if the machine learning and data analysis algorithms are handled carefully.}.  It employs one motor to allow it to rotate about a vertical axis indicated by the black line in top center of the figure, and two motors to extend and lower the arm about the joints located at the positions indicated by the short arrows.  The motors are controlled directly by the LEGO brick, which is commanded via Bluetooth by a Dell Latitude laptop computer running the robot's machine learning system, which is programmed in MatLab (Mathworks, Inc.).  The LEGO light sensor is attached to the end of the arm (indicated by the long arrow in Figure \ref{fig:arm-sensor} and displayed in the insight at the upper right).  Based on commands issued by the laptop, the robotic arm can deploy the light sensor to any position within its reach on the playing field.  The light sensor is lowered to an average height of 14mm above the surface before taking a measurement.  The arm is designed using a trapezoidal construction that maintains the sensor's orientation to be aimed at nadir, always normal to the surface despite the extension of the arm.

The LEGO light sensor (LEGO Part 9844) consists of a photodiode-LED pair.  The white circle is the photo diode, and the red circle is the illuminating LED. Note that they are separated by a narrow plastic ridge, which prevents the LED from shining directly into the photo diode. This ridge, along with the plastic lenses and the presence of the illuminating LED, each affect the spatial sensitivity of the sensor.  When activated, the light sensor   flashes for a brief instant and measures the intensity of the reflected light.  The photo diode and its support circuitry are connected to the sensor port of the LEGO Brick (LEGO Part 9841), which runs on a 32-bit ARM7 ATMEL micro-controller.
The measured intensities are converted by internal software running on the ATMEL micro-controller to a scale of $1$ to $100$, which we refer to as \emph{LEGO units}.  The light sensor integrates the light arriving from a spatially distributed region within its field of view.  The spatial sensitivity of the sensor to light sources within its field of view is described by the Spatial Sensitivity Function (SSF).  This function is unknown, but if it were known, one could weight the surface albedo field with the SSF and integrate to obtain an estimate of the recorded intensity of the reflected light.

\subsection{The Circle Characterization Experiment}
The robotic arm is designed to deploy the light sensor to take measurements of the surface albedo at locations within a playing field of dimensions approximately $131 \times 65$ LEGO distance units (1048mm x 520mm), within an automated experimental design paradigm \cite{knuth_arm_2007,knuth_center2008,knuth_center2010}.  The robot is programmed with a set of hypotheses of what shapes it could find placed on the playing field.  Instead of being programmed with a specific set of strategies for characterizing the hypothesized shapes, the robot utilizes a generalized Bayesian Inference Engine coupled to an Inquiry Engine to make inferences about the hypotheses based on the recorded data and to use uncertainties in its inferences to drive further exploration by autonomously selecting new measurement location that promise to provide the maximum amount of relevant information about the problem.

In this experiment, the robotic arm is instructed that there is a white circle of unknown radius arbitrarily placed on the black field.  Such an instruction is encoded by providing the robot with a model of the surface albedo consisting of three parameters: the center location $(x_o, y_o)$ and radius $r_o$, written jointly as
\begin{equation} \label{eq:circle_model}
\mathbf{C} = \{(x_o, y_o), r_o\},
\end{equation}
so that given a measurement location $(x_i, y_i)$ the albedo $S$ is expected to be
\begin{equation}
S(x_i, y_i ; \mathbf{C}) = \\
 \begin{cases}
 1 & \text{if $D((x_i, y_i),(x_o, y_o)) \leq r_o$}\\
 0 & \text{if $D((x_i, y_i),(x_o, y_o)) > r_o$},
\end{cases}
\end{equation}
where
\begin{equation} \label{eq:distance}
D((x_i, y_i),(x_o, y_o)) = \sqrt{(x_i-x_o)^2 + (y_i-y_o)^2}
\end{equation}
is the Euclidean distance between the measurement location $(x_i, y_i)$ and the center of the circle $(x_o, y_o)$, and an albedo of $1$ signifies that the surface is white, and 0 signifies that the surface is black.
Precisely how these expectations are used to make inferences from data is explained in the next section.  Keep in mind that while the circle's precise radius and position is unknown, the robot has been provided with limited prior information about the allowable range of radii and positions.

Again, it is important to note that the robot does not scan the surface to solve the problem, nor does it try to find three points along the edge of the circle.  Instead, it employs a general system that works for any expected shape or set of shapes that autonomously and intelligently determines optimal measurement locations based both on what it knows and on what it does not know.  The number of measurements needed to characterize all three circle parameters to within the desired accuracy is a measure of the efficiency of the experimental procedure.

\subsection{The Machine Learning System}
The machine learning system employs a Bayesian Inference Engine to make inferences about the circle parameters given the recorded light intensities, as well as an Inquiry Engine designed to use the uncertainties in the circle parameter estimates to autonomously select measurement locations that promise to provide the maximum amount of relevant information about the problem.

The core of the Bayesian Inference Engine is centered around the computation of the \emph{posterior probability} $Pr(\mathbf{C} | \mathbf{D}, I)$ of the albedo model parameters, $\mathbf{C}$ in (\ref{eq:circle_model}) given the light sensor recordings (data) $d_i$ recorded at locations $(x_i, y_i)$, which we write compactly as
\begin{equation}
\mathbf{D} = \{(d_1, (x_1, y_1)), \ldots, (d_N, (x_N, y_N))\},
\end{equation}
and any additional prior information $I$.
Bayes' Theorem allows one to write the posterior probability as a function of three related probabilities
\begin{equation}
Pr(\mathbf{C} | \mathbf{D}, I) = Pr(\mathbf{C} | I) \frac{Pr(\mathbf{D} | \mathbf{C}, I)}{Pr(\mathbf{D} | I)},
\end{equation}
where the right-hand side consists of the product of the \emph{prior probability} of the circle parameters $Pr(\mathbf{C} | I)$, which describes what is known about the circle before any sensor data are considered, with a ratio of probabilities that are sensor data dependent.  It is in this sense that Bayes' Theorem represents a learning algorithm since what is known about the circle parameters before the data are considered (prior probability) is modified by the recorded sensor data resulting in a quantification of what is known about the circle parameters after the data are considered (posterior probability).  The probability in the numerator on the right is the \emph{likelihood} $Pr(\mathbf{D} | \mathbf{C}, I)$, which quantifies the probability that the sensor data could have resulted from the hypothesized circle parameterized by $\mathbf{C}$.  The probability in the denominator is the \emph{marginal likelihood} or the \emph{evidence}, which here acts as a normalization factor.  Later when estimating the SSF of the light sensor (which is a different problem), the evidence, which can be written as the integral of the product of the prior and the likelihood over all possible model parameters, will play a critical role.

The likelihood term, $Pr(\mathbf{D} | \mathbf{C}, I)$, plays a critical role in the inference problem, since it essentially compares predictions made using the hypothesized circle parameters to the observed data.  A na\"{i}ve light sensor model would predict that if the sensor was centered on a black region, it would return a small number, and if the sensor was centered on a white region, it would return a large number.  A more accurate light sensor model would take into account the SSF of the light sensor and perform an SSF-weighted integral of the hypothesized albedo field and compare this to the recorded sensor data.  These two models of the light sensor will be discussed in detail in the next section.

 \begin{figure}[tb]  \centering
 \includegraphics[width=0.90\textwidth]{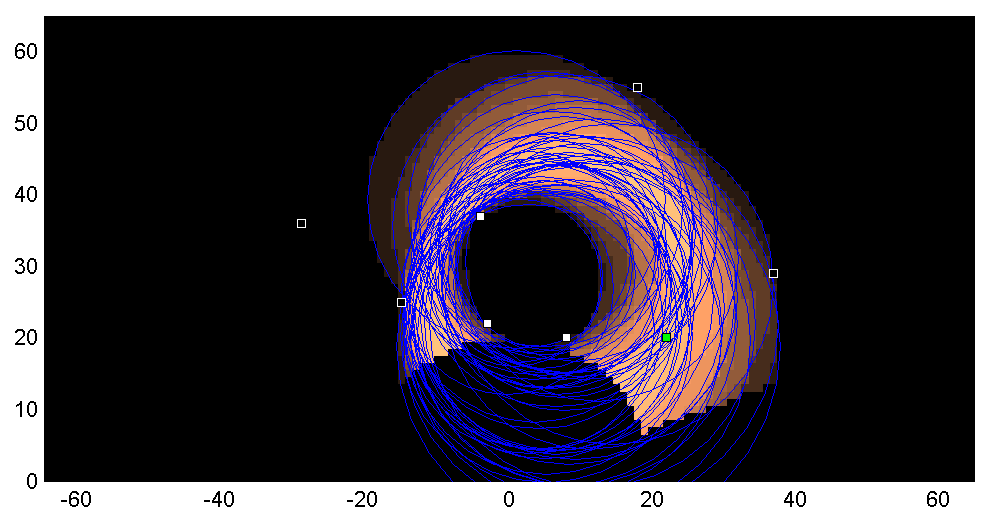}
\caption{This figure illustrates the robot's machine learning system's view of the playing field using the na\"{i}ve light sensor model.  The axes label playing field coordinates in LEGO distance units.  The previously obtained measurement locations used to obtain light sensor data are indicated by the black and white squares indicating the relative intensity with respect to the na\"{i}ve light sensor model.  The next selected measurement location is indicated by the green square.  The blue circles represent the 50 hypothesized circles sampled from the posterior probability.  The shaded background represents the entropy map, such that brighter areas indicate the measurement locations that promise to provide greater information about the circle to be characterized.  Note that the low entropy area bounded by the white squares indicates that this region is probably inside the white circle and that measurements made here will not be as informative as measurements made elsewhere.  The dark jagged edges at the bottom of colored high entropy region reflect the boundary between the playing field and the region that is outside of the robotic arm's reach.}
\label{fig:entropy-map}
\end{figure}

The robot not only makes inferences from data, but it designs its own experiments by autonomously deciding where to take subsequent measurements \cite{knuth_intelligent_2003,knuth_arm_2007,knuth_center2008, knuth_center2010,malakar_entropy-based_2010,malakar_2011}.  This can be viewed in terms of Bayesian experiment design \cite{ lindley_measure_1956, fedorov1972theory, Chaloner95bayesianexperimental, sebastiani_bayesian_1997, sebastiani2000maximum, Loredo03bayesianadaptive, fischer2004bayesian} where the Shannon entropy \cite{shannon_mathematical_1949} is employed as the utility function used to decide where to take the next measurement.  In short, the Bayesian Inference Engine samples 50 circles from the posterior probability using the nested sampling algorithm \cite{skilling_nested_2004,sivia_data_2006}.  Since these samples may consist of replications, they are further diversified by taking approximately 1000 Metropolis-Hastings steps \cite{Metropolis:1953,Hastings:1970}.  Figure \ref{fig:entropy-map} illustrates the robot's machine learning system's view of the playing field after several measurements have been recorded.  The 50 circles sampled from the posterior probability (blue circles) reflect what the robot knows about the white circle (not shown) given the light sensor data that it has collected.  The algorithm then considers a fine grid of potential measurement locations on the playing field.  At each potential measurement location, the 50 sampled circles are queried using the likelihood function to produce a sample of what measurement could be expected at that location given each hypothesized circle.  This results in a set of potential measurement values at each measurement location.  The Shannon entropy of the set of potential measurements at each location is computed, which results in an \emph{entropy map}, which is illustrated in Figure \ref{fig:entropy-map} as the copper-toned coloration upon which the blue circles lie.  The set of measurement locations with the greatest entropy are identified, and the next measurement location is randomly chosen from that set.  Thus the likelihood function not only affects the inferences about the circles made by the machine learning algorithm, but it also affects the entropy of the measurement locations, which guides further exploration.  In the next section we describe the two models of the light sensor and their corresponding likelihood functions.

The efficacy of the sensor model will be quantified by the average number of measurements the robot needs to make to estimate the circle parameters within a given precision.

\subsection{Models Describing a Light Sensor}
In this section we discuss two models of a light sensor and indicate precisely how they are integrated into the likelihood function used by both the Bayesian Inference and Inquiry Engines.

\subsubsection{The Na\"{i}ve Likelihood}
A na\"{i}ve light sensor model would predict that if the sensor was centered on a black region (surface albedo of zero), the sensor would return a small number on average, and if it were centered on a white region (surface albedo of unity), it would return a large number on average.  Of course, there are expected to be noise variations from numerous sources, such as uneven lighting of the surface, minor variations in albedo, and noise in the sensor itself.  So one might expect that there is some expected squared deviation from the two mean sensor values for the white and black cases.  For this reason, we model the expected sensor response with a Gaussian distribution with mean $\mu_B$ and standard deviation $\sigma_B$ for the black surface, and a Gaussian distribution with mean $\mu_W$ and standard deviation $\sigma_W$ for the white surface.  The likelihood of a measurement $d_i$ at location $(x_i, y_i)$ corresponding to the na\"{i}ve light sensor model, $Pr_{naive}(\{(d_i, (x_i, y_i))\} | \mathbf{C}, I)$, can be written compactly as:
\begin{equation}
Pr_{naive}(\{(d_i, (x_i, y_i))\} | \mathbf{C}, I) =
 \begin{cases}
 (2\pi \sigma_W)^{-1/2} \exp{\Big[\frac{(\mu_W-d_i)^2}{2 \sigma_W^2}\Big]} & \text{for $D((x_i, y_i),(x_o, y_o)) \leq r_o$}\\
 (2\pi \sigma_B)^{-1/2} \exp{\Big[\frac{(\mu_B-d_i)^2}{2 \sigma_B^2}\Big]} & \text{for $D((x_i, y_i),(x_o, y_o)) > r_o$},
\end{cases}
\end{equation}
where $\mathbf{C} = \{(x_o, y_o), r_o\}$ represents the parameters of the hypothesized circle and $D((x_i, y_i),(x_o, y_o))$ is the Euclidean distance given in (\ref{eq:distance}).  The joint likelihood for $N$ independent measurements is found by taking the product of the $N$ single-measurement likelihoods.
In a practical experiment, the means $\mu_B$ and $\mu_W$ and standard deviations $\sigma_B$ and $\sigma_W$ can be easily estimated by sampling known black and white regions several times using the light sensor.

\subsection{The SSF Likelihood}\label{sec:SSF-Likelihood}
A more accurate likelihood can be developed by taking into account the fact that the photo diode performs a weighted-integral of the light arriving from a spatially distributed region within its field of view, the weights being described by the Spatial Sensitivity Function (SSF) of the sensor.  Since the SSF of the light sensor could be arbitrarily complex with many local peaks, but is expected to decrease to zero far from the sensor, we characterize it using a mixture of Gaussians (MoG) model, which we describe in this section.

The sensor's response situated a fixed distance above the point $(x_i, y_i)$ to a known black-and-white pattern can be modeled in the lab frame by
\begin{equation}
\label{eq:modelI}
M(x_i, y_i) = I_{min} + \left( I_{max} - I_{min}\right) R\left(x_i, y_i\right),
\end{equation}
where $I_{min}$ and $I_{max}$ are observed intensities for a completely black surface (surface albedo of zero) and a completely white surface (surface albedo of unity), respectively, and $R$ is a scalar response function, varying between zero and one, that depends both on the SSF and the surface albedo \cite{malakar_SSF2009}.  The minimum intensity $I_{min}$ acts as an offset, and $(I_{max} - I_{min})$ serves to scale the response to LEGO units.


The response of the sensor is described by $R(x_i, y_i)$, which is a convolution of the sensor SSF and the surface albedo $S(x,y)$ given by
\begin{equation}
\label{eq:RwithSSF}
R(x_i,y_i) = \int dx ~dy ~SSF(x-x_i , y-y_i) ~S(x,y).
\end{equation}
where the SSF is defined so that the convolution with a completely white surface results in a response of unity.  In practice, we approximate this integral as a sum over a pixelated grid with 1mm square pixels
\begin{equation}
\label{eq:discreteR}
R(x_i,y_i) = \sum_{x,y}~SSF(x-x_i , y-y_i)~S(x,y).
\end{equation}

We employ a mixture of Gaussians (MOG) as a parameterized model to describe the SSF in the sensor's frame co-ordinates $(x', y') = (x-x_i , y-y_i)$
\begin{equation}
\begin{split}
\label{Eq:ssf}
SSF(x',y')
  = \frac{1}{K} & \sum_{n=1}^{N} a_n ~\times\\
  &   \exp\left[ -\left\{A_n (x' - u'_n)^2+ B_k (y' -v'_n)^2+ 2 C_n (x'-u'_n)(y'-v'_n)\right\}\right],
\end{split}
\end{equation}
where $(u'_n, v'_n)$ denotes the center of the $n^{th}$ two-dimensional Gaussian with amplitude $a_n$ and covariance matrix elements given by $A_n, B_n $ and $C_n$. The constant $K$ denotes a normalization factor, which ensures that the SSF integrates to unity \cite{malakar_SSF2009}. The model is sufficiently general so that one could vary the number of Gaussians to any number, although we found that in modeling this light sensor testing models with $N$ varying from $N=1$ to $N=4$ was sufficient.
The MOG model results in a set of six parameters to be estimated for each Gaussian
$$ \theta_n = \left\{a_n , u'_n, v'_n, A_n, B_n, C_n \right\},$$ where the subscript $n$ is used to denote the $n^{th}$ Gaussian in the set.  These must be estimated along with the two intensity parameters, $I_{min}$ and $I_{max}$ in (\ref{eq:modelI}) so that an MOG model with $N$  Gaussians consists of $6 N + 2$ parameters to be estimated.

We assign a Student-t distribution to the SSF model likelihood, which can be arrived at by assigning a Gaussian likelihood and integrating over the unknown standard deviation $\sigma$ \cite{sivia_data_2006}. By defining $\mathbf{D} = \{(d_1,(x_1,y_1)), (d_2,(x_2,y_2)), \ldots, (d_N,(x_N,y_N))\}$ and writing the hypothesized circle parameters as $\mathbf{C} = \{(x_o, y_o), r_o\}$, we have
\begin{equation}
Pr_{SSF}(\mathbf{D} | \mathbf{C}, I) \propto \left[ \sum_{i=1}^{N}\left( M(x_i, y_i) - d_i \right)^2 \right] ^{-(N-1)/2} .
\end{equation}
where $N$ is the number of measurements made by the robot, and the function $M$, defined in (\ref{eq:modelI}), relies on both (\ref{eq:RwithSSF}) and (\ref{Eq:ssf}).  Note that in practice, the MoG SSF model described in (\ref{Eq:ssf}) is used to generate a discrete SSF matrix which is used in the convolution (\ref{eq:discreteR}) to compute the likelihood function via the sensor response model $M$ in (\ref{eq:modelI}).

\subsection{Data Collection for SSF Estimation}
In this section we describe the collection of the light sensor readings in the laboratory that were used to estimate the SSF of the light sensor.  The SSF is not only a function of the properties of the photo diode, but also of the illuminating LED and the height above the surface.  For this reason, measurements were made at a height of 14mm above a surface with a known albedo in a darkened room to avoid complications due to ambient light and to eliminate the possibility of shadows cast by the sensor or other equipment \cite{malakar_SSF2009}.

\begin{figure}[tb]
\centering	\includegraphics[width=0.40\textwidth]{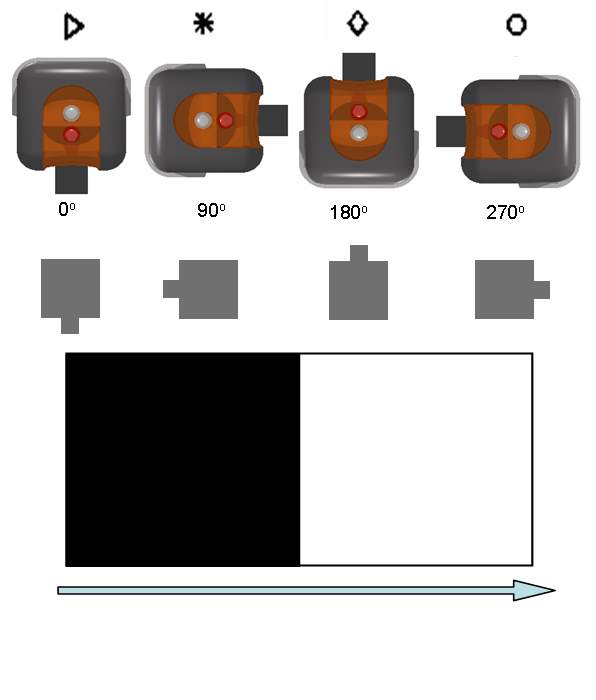}	 \vspace{-10pt}
\caption[Sensor Orientations]{(Bottom) This figure illustrates the laboratory surface, referred to as the \emph{black-and-white boundary pattern}, with known albedo, which consisted of two regions: a black region on the left and a white region on the right separated by a sharp linear boundary. (Top) This illustrates the four sensor orientations used to collect data for the estimation of the SSF along with the symbols used to indicate the measured values plotted in Figure \ref{fig:data}.  The top row views the sensors as if looking up from the table so that the photodiode placement in the sensor package can be visualized.  Below these the gray sensor shapes illustrate how they are oriented looking down at both the sensors and the albedo surface.  Measurements were taken as the sensor was incrementally moved, with one millimeter steps, in a direction perpendicular to the boundary (as indicated by the arrow at the bottom of the figure) from a position of 5cm to the left of the boundary (well within the completely black region) to a position 5cm to the right of the boundary (well within the completely white region).  This process was repeated four times with the sensor in each of the four orientations.}
	\label{fig:orientation}
\end{figure}

The surface, which we refer to as the \emph{black-and-white boundary pattern}, consisted of two regions: a black region on the left and a white region on the right separated by a sharp linear boundary as shown in Figure \ref{fig:orientation} (bottom).  Here the surface and sensor are illustrated as if viewing them by looking up at them from below the table surface.  This is so the placement of the photodiode in the sensor package can be visualized.
The lab frame was defined to be at the center of the black-white boundary so that the surface albedo, $S_{BW}(x,y)$, is given by
\begin{equation}
 S_{BW}(x,y) =
 \begin{cases}
 1, & \text{for $ x > 0$}\\
 0, & \text{for $x \le 0$}
 \\
\end{cases}
\end{equation}
Measurements were taken as the sensor was incrementally moved, with one millimeter steps, in a direction perpendicular to the boundary from a position of 5cm to the left of the boundary (well within the completely black region) to a position 5cm to the right of the boundary (well within the completely white region) resulting in 101 measurements.  This process was repeated four times with the sensor in each of the four orientations (see Figure \ref{fig:orientation} for an explanation), giving a total of 404 measurements using this pattern.

\begin{figure}[tb]
    \centering
    \includegraphics[width=0.30\textwidth]{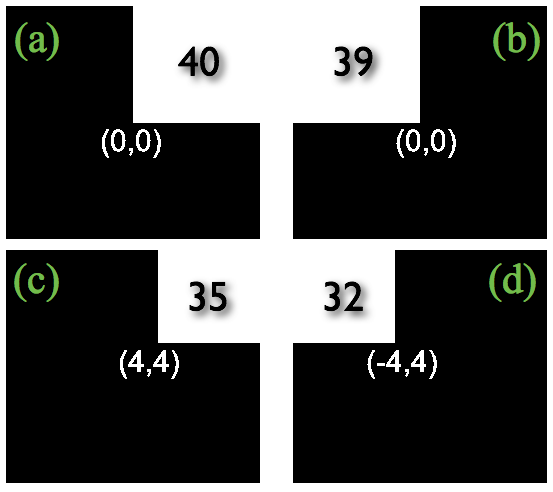}
    \caption{Four additional symmetry-breaking albedo patterns were employed.  In all cases, the sensor was placed in the $0^o$ orientation at the center of the pattern, indicated by $(0,0)$.  In the two lower patterns, the center of the white square was shifted diagonally from the center by 4mm as indicated by the coordinates (white text) of the corner.  The recorded intensity levels are displayed in the white albedo area.}	\label{four-extra}
\end{figure}

The black-and-white boundary pattern does not provide sufficient information to uniquely infer the SSF, since the sensor may have a response that is symmetric about a line oriented at $45^o$ with respect to the linear boundary.  For this reason, we employed four additional albedo patterns consisting of black regions with one white quadrant as illustrated in Figure \ref{four-extra}, which resulted in four more measurements.
These have surface albedos defined by
\begin{eqnarray} \label{eq:additional_albedos}
S_a(x,y) & =
 \begin{cases}
 1, & \text{for $ x > 0mm$ and $y > 0mm$}\\
 0, & \text{otherwise}
 \\
\end{cases}\\
 S_b(x,y) & =
 \begin{cases}
 1, & \text{for $ x <  0mm$ and  $y > 0mm$}\\
 0, & \text{otherwise}
 \\
\end{cases}\\
 S_c(x,y) & =
 \begin{cases}
 1, & \text{for $ x > 4mm$ and $y > 4mm$}\\
 0, & \text{otherwise}
 \\
\end{cases}\\
 S_d(x,y) & =
 \begin{cases}
 1, & \text{for $ x < -4mm$ and $y > 4mm$}\\
 0, & \text{otherwise}
 \\
\end{cases}
\end{eqnarray}
where the subscripts relate each albedo function to the pattern illustrated in Figure \ref{four-extra}.

\subsection{Estimating SSF MoG Model Parameters}\label{sec:estimating-SSF}
In this section we describe the application of Bayesian methods to estimate the SSF MoG model parameters.  Keep in mind that in this paper we are considering two distinct inference problems: the robot's inferences about a circle and our inferences about the light sensor SSF.  Both of these problems rely on making predictions about measured intensities using a light sensor.  For this reason many of these equations will not only look similar to what we have presented above, but also depend on the same functions mapping modeled albedo fields to predicted sensor responses, which are collectively represented using the generic symbol $\mathbf{D}$.  It may help to keep in mind that the robot is characterizing a circle quantified by model parameters represented jointly by the symbol $\mathbf{C}$ and we are estimating the SSF of a light sensor quantified by model parameters represented jointly by the symbol $\theta$ below.  Aside from the evidence, represented by function $Z$ below (which is not used by the robot in this experiment), all of the probability functions contain the model parameters in their list of arguments making it clear to which inference problem they refer.

The posterior probability for the SSF MoG model parameters, collectively referred to as $\theta = \{\theta_1, \theta_2, \ldots, \theta_N\}$, for a model consisting of $N$ Gaussians is given by
\begin{equation}
\label{eq:bayes}
Pr(\theta|\textbf{D}, I) =  \frac{1}{Z} ~Pr(\theta | I) ~Pr(\textbf{D} |\theta, I),
\end{equation}
where here $\mathbf{D}$ refers to the data collected for the SSF estimation experiment described in the previous section, $I$ refers to our prior information about the SSF (which is that it may have several local peaks and falls off to zero far from the sensor), and $Z$ refers to the evidence $Z = Pr(\textbf{D} |I)$, which can be found by
\begin{equation}
\label{Evidence1}
 Z = \int d\theta Pr(\theta | I) Pr(\textbf{D} |\theta, I).
\end{equation}
In our earlier discussion where Bayes' theorem was used to make inferences about circles, the evidence played the role of a normalization factor.  Here, since we can consider MoG models with different numbers of Gaussians and since we integrate over all of the possible values of the parameters, the evidence quantifies the degree to which the hypothesized model order $N$ supports the data.  That is, the optimal number of Gaussians to be used in the MoG model of the SSF can be found by computing the evidence.

All five sets of data $\mathbf{D}$ described in the previous section were used to compute the posterior probability.  These are each indexed by the subscript $i$ where $i = 1,2,3,4$, so that $D_i$ refers to the data collected using each of the four orientations $\phi_i = \left\{0^o, 90^o, 180^o, 270^o \right\}$ to scan the black-and-white boundary pattern resulting in $N_i = 101$ measurements for $i = 1,2,3,4$, and where the value $i=5$ refers to the $N_i = 4$ measurements attained using the $0^o$ orientation with the set of four additional patterns.

We assign uniform priors so that this is essentially a maximum likelihood calculation with the posterior being proportional to the likelihood.  We assign a Student-t distribution to the likelihood, which can be arrived at by assigning a Gaussian likelihood and integrating over the unknown standard deviation $\sigma$ \cite{sivia_data_2006}. This can be written as
\begin{equation}
 \label{eq:studentT}
 Pr(D_i | ~\theta, I) \propto \left[ \sum_{j=1}^{N_i} \left( M_{ij}(x_{ij},y_{ij}) - D_i(x_{ij}, y_{ij}) \right)^2 \right] ^{-(N_i-1)/2} .
\end{equation}
where $i$ denotes each of the five sets of data and $j$ denotes the $j^{th}$ measurement of that set, which was taken at position $(x_{ij},y_{ij})$.  The function $M_{ij}(x,y)$ represents a predicted measurement value obtained from (\ref{eq:modelI}) using (\ref{eq:discreteR}) with the albedo function $S(x,y)$ defined using the albedo pattern $S_{BW}(x,y)$ with orientation $\phi_i$ for $i = 1,2,3,4$ and the albedo patterns $S_a, S_b, S_c, S_d$ for $i=5$ and $j=1,2,3,4$, respectively.  As such the likelihood relies on a difference between predicted and measured sensor responses.

The joint likelihood for the five data sets is found by taking the product of the likelihoods for each data set, since we expect that the standard deviations that were marginalized over to get the Student-t distribution could have been different for each of the five data sets as they were not all recorded at the same time
\begin{equation}
Pr(\textbf{D} | ~\theta, I) = \prod_{i=1}^{5} Pr(D_i| ~\theta, I).
\end{equation}

We employed nested sampling \cite{skilling_nested_2004,sivia_data_2006} to explore the posterior probability since, in addition to providing parameter estimates, it is explicitly designed to perform evidence calculations, which we use to perform model comparison in identifying the most probable number of Gaussians in the MoG model.  For each of the four MoG models (number of Gaussians varying from one to four) the nested sampling algorithm was initialized with 300 samples and iterated until the change in consecutive log-evidence values less than $10^{-8}$.  Typically one estimates the mean parameter values by taking an average of the samples weighted by a quantity computed by the nested samp[ling algorithm called the logWt in the references \cite{skilling_nested_2004,sivia_data_2006}.  Here we simply performed a logWt-weighted average of the sampled SSF fields computed using the sampled MoG model parameters (rather than the logWt-weighted average of the MoG model parameters themselves), so the result obtained using an MoG model consisting of a single Gaussian is not strictly a single two-dimensional Gaussian distribution.  It is this discretized estimated SSF field matrix that is used directly in the convolution (\ref{eq:discreteR}) to compute the likelihood functions as mentioned earlier in the last lines of Section \ref{sec:SSF-Likelihood}.

\section{Results and Discussion}
In this section we present the SSF MoG light sensor model estimated from the laboratory data, and evaluate its efficacy by demonstrating a significant improvement of the performance in the autonomous robotic platform.

\subsection{Light Sensor SSF MoG Model}

\begin{figure}[tb]
\centering	\includegraphics[width=0.70\textwidth]{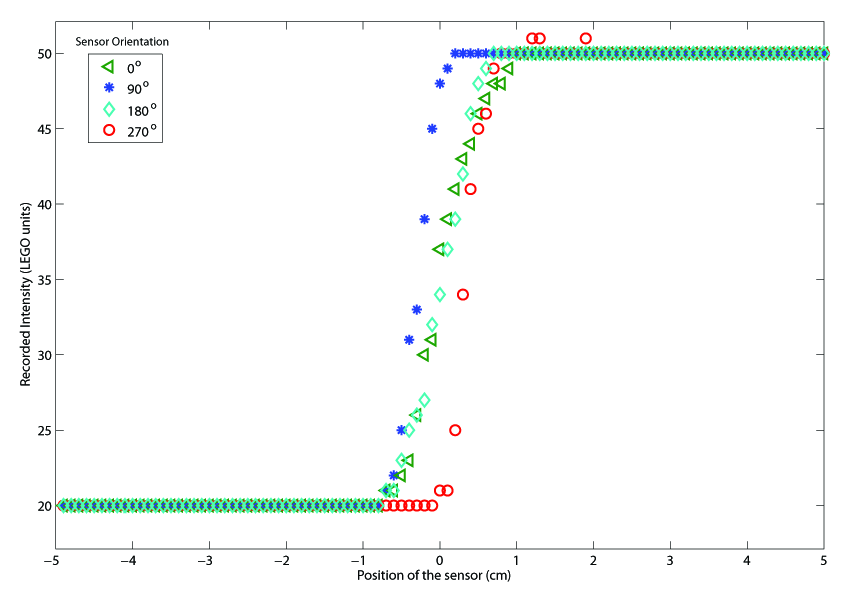}
    \caption{This figure illustrates the intensity measurements, $D_1, D_2, D_3, D_4$ from the four sensor orientations, recorded from the sensor using the black-and-white boundary pattern.  Figure \ref{fig:orientation} shows the orientations of the sensor corresponding to the symbols used this figure.}
	\label{fig:data}
\end{figure}

The light sensor data collected using the black-and-white boundary pattern are illustrated in Figure \ref{fig:data}.  One can see that the intensity recorded by the sensor increases dramatically as the sensor crosses the boundary from the black region to the white region, but that the change is not a step function, which indicates the finite size of the surface area integrated by the sensor.  It is this effect that is to be modeled by the SSF function.  There is an obvious asymmetry between the $90^o$ and $270^o$ orientations due to the shift of the transition region.  In addition, there is a significant different in slope of the transition region between the $0^o,180^o$ orientations and the $90^o,270^o$ orientations indicating a significant difference in the width of the SSF in those directions.  Note also that the minimum recorded response is not zero, as the reflectance of the black surface is not completely zero.

 \begin{figure}[tb]
\centering
\includegraphics[width=0.6\textwidth]{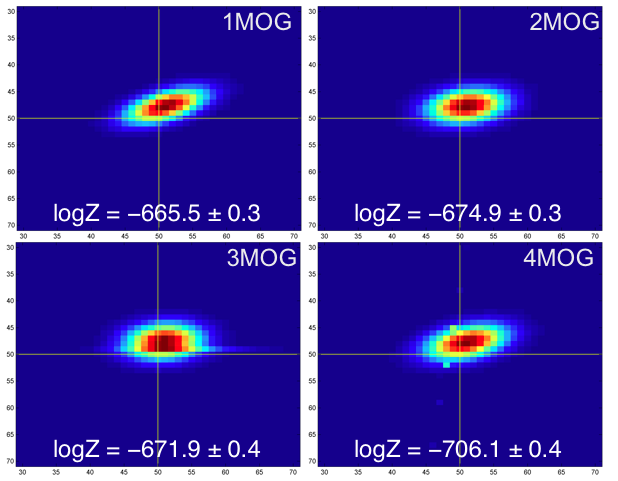}
\caption{This figure illustrates the SSF obtained from the four MOG models along with their corresponding log-evidence values.}
\label{fig:mogSSF}
\end{figure}
\begin{table}\caption{A comparison of the tested MoG SSF Models and their respective Log-Evidence (in units of $data^{-408}$).} \label{tableMoG}
 \centering
    \begin{tabular}{cccc}
    \hline
    \hline
    MoG Model Order & Log-Evidence &  Number of Parameters \\
    \hline
    1 Gaussian & $-665.5\pm 0.3$ & 6 \\
    2 Gaussian & $-674.9\pm 0.3$ & 12 \\
    3 Gaussian & $-671.9\pm 0.4$ & 18 \\
    4 Gaussian & $-706.1\pm 0.4$ & 24 \\
    \hline
    \end{tabular}
    \end{table}

\begin{figure}[tb]
\centering
\includegraphics[width=0.7\textwidth]{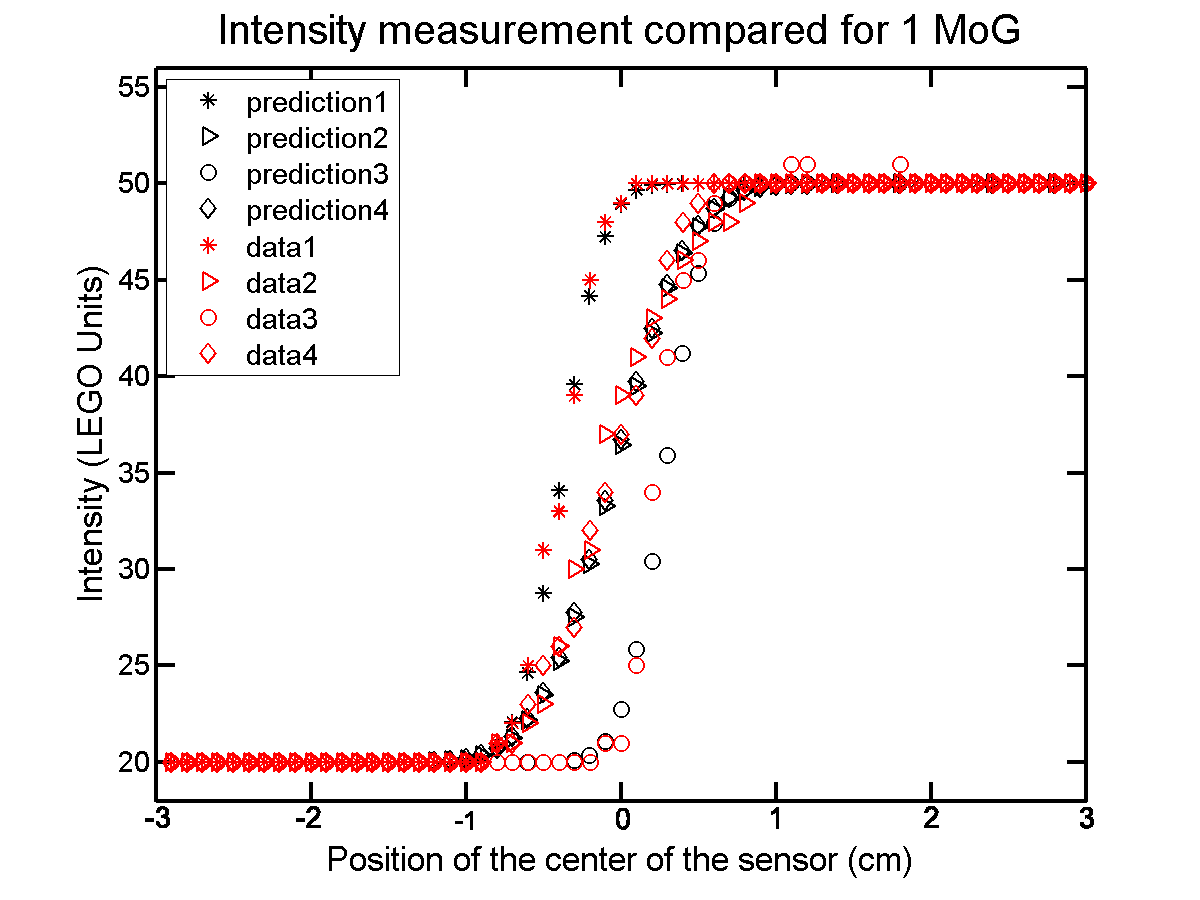}
\caption{A comparison of the observed data (red) with predictions (black) made by the SSF field estimated using the single two-dimensional Gaussian MoG model. } \label{fig:1mogprediction}
\end{figure}

The nested sampling algorithm produced the mean SSF fields for each of the MoG models tested, as well as the corresponding log-evidence.  Table \ref{tableMoG}, which lists the log-evidence computed for each MoG model order, illustrates that the most probable model was obtained from the single two-dimensional Gaussian models by a factor of about $\exp(9)$, which means that it is about 8000 times more probable than the MoG consisting of two Gaussians.  Figure \ref{fig:mogSSF} shows the mean SSF fields described by the MOG models of different orders.  In all cases, the center of the SSF is shifted slightly above the physical center of the sensor package due to the placement of the photodiode (refer to Figure \ref{fig:arm-sensor}) as indicated by the data in Figure \ref{fig:data}.  In addition, as predicted, one sees that the SSF is wider along the $90^o-270^o$ axis than along the $0^o-180^o$ axis.  Last, Figure \ref{fig:1mogprediction} demonstrates that the predicted light sensor output shows excellent similarity to the recorded data for the black-and-white boundary pattern.

In the next section, we demonstrate that explicit knowledge about how the light sensor integrates light arriving from within its field-of-view improves the inferences one can make from its output.

\subsection{Efficacy of Sensor Model}
The mean SSF field obtained using a single two-dimensional Gaussian model (Figure \ref{fig:mogSSF}, upper left), was incorporated into the likelihood function used by the robot's machine learning system.  Here we compare the robot's performance in locating and characterizing a circle by observing the average number of measurements necessary to characterize the circle parameters to within a precision of 4mm (which is one-half of the spacing between the holes in the LEGO Technic Parts).

 \begin{figure}[p]  \centering
 \includegraphics[width=0.90\textwidth]{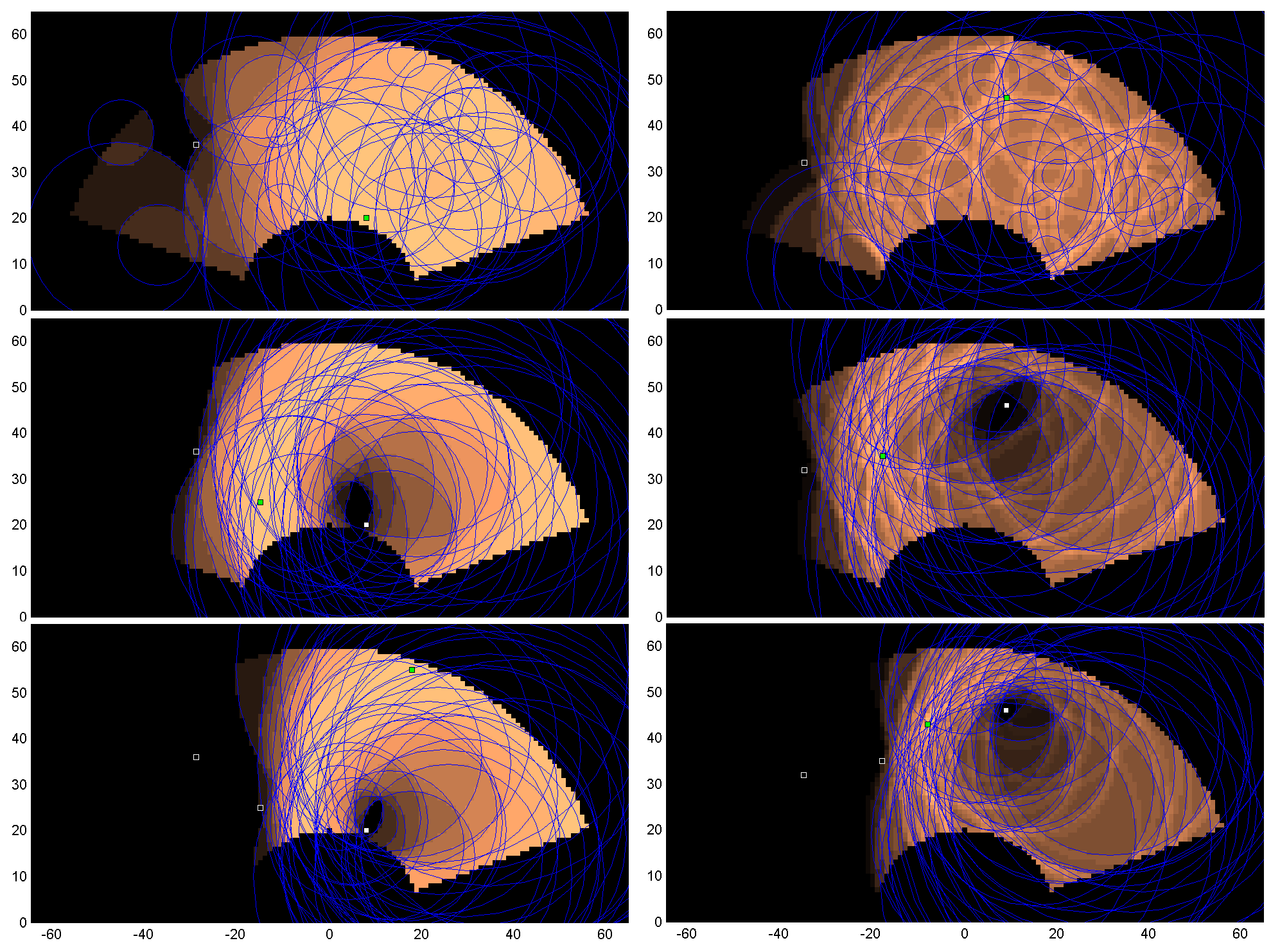}
\caption{(Left Column) These three panels illustrate the robot's machine learning system's view of the playing field using the na\"{i}ve light sensor model as the system progresses through the first three measurements.  The previously obtained measurement locations used to obtain light sensor data are indicated by the black and white squares indicating the relative intensity with respect to the na\"{i}ve light sensor model.  The next selected measurement location is indicated by the green square.  The blue circles represent the 50 hypothesized circles sampled from the posterior probability.  The shaded background represents the entropy map, which indicates the measurement locations that promise to provide maximal information about the circle to be characterized.  Note that the low entropy area surrounding the white square indicates that the region is probably inside the white circle (not shown) and that measurements made there will not be as informative as measurements made elsewhere.  The entropy map in Figure \ref{fig:entropy-map} shows the same experiment at a later stage after seven measurements have been recorded.  (Right Column) These three panels illustrate the robot's machine learning system's view of the playing field using the more accurate SSF light sensor model.  Note that the entropy map reveals the circle edges to be highly informative. This is because it helps not only to identify whether the sensor is inside the circle (as is accomplished using the na\"{i}ve light sensor model on the left), but also the extent to which the sensor is on the edge of the circle.}
\label{fig:entropy-map-scenes}
\end{figure}

The three panels comprising the left column of Figure \ref{fig:entropy-map-scenes} illustrate the robot's machine learning system's view of the playing field using the na\"{i}ve light sensor model.  The previously obtained measurement locations used to obtain light sensor data are indicated by the black and white squares indicating the relative intensity with respect to the na\"{i}ve light sensor model.  The next selected measurement location is indicated by the green square.  The blue circles represent the 50 hypothesized circles sampled from the posterior probability.  The shaded background represents the entropy map, which indicates the measurement locations that promise to provide maximal information about the circle to be characterized.  The low entropy area surrounding the white square indicates that the region is probably inside the white circle (not shown) and that measurements made there will not be as informative as measurements made elsewhere.  Note that the circles partition the plane, and that each partition has a uniform entropy.  All measurement locations within that partition, or any other partition sharing the same entropy, all stand to be equally informative.  Here it is the fact that the shape is known to be a circle that is driving the likelihood function.

In contrast, the three panels comprising the right column of Figure \ref{fig:entropy-map-scenes} illustrate the playing field using the more accurate SSF model.  Here one can see that the entropy is higher along the edges of the sampled circles.  This indicates that the circle edges promise to provide more information than the centers of the partitioned regions.  This is because the SSF model enables one to detect not only whether the light sensor is situated above the circles edge but also how much of the SSF overlaps with the white circle.  That is, it helps not only to identify whether the sensor is inside the circle (as is accomplished using the na\"{i}ve light sensor model), but also the extent to which the sensor is on the edge of the circle.  The additional information provided about the functioning of the light sensor translates directly into additional information about the albedo that results in the sensor output.

This additional information can be quantified by observing how many measurements the robot is required to take to obtain estimates of the circle parameters to within the same precision in the cases of each light sensor model.  Our experiments revealed that on average it takes $26\pm$ measurements using the na\"{i}ve light sensor model compared to an average of $16\pm$ measurements for the more accurate SSF light sensor model.

\section{Conclusion}
The quality of the inferences one makes from a sensor depend not only on the quality of the data returned by the sensor, but also on the information one possesses about the sensor's performance.  In this paper we have demonstrated via a case study, how more precisely modeling a sensor's performance can improve the inferences one can make from its data.  In this case, we demonstrated that one can achieve an almost 40\% reduction of the number of measurements needed by a robot to make the same inferences by more precisely modeling its light sensor.

This paper demonstrates how a machine learning system that employs Bayesian inference (and inquiry) relies on the likelihood function of the data given the hypothesized model parameters.  Rather than simply representing a noise model, the likelihood function quantifies the probability that a hypothesized situation could have given rise to the recorded data.  By incorporating more information about the sensors (or equipment) used to record the data, one naturally is incorporating this information into the posterior probability, which results in one's inferences.

This is made even more apparent by a careful study of the experimental design problem that this particular robotic system is designed to explore.  For example, it is easy to show that by using the na\"{i}ve light sensor model, the entropy distribution for a proposed measurement location depends solely on the number of sampled circles for which the location is in the interior of the circle, and the number of sampled circles for which the location is exterior to the circle.  Given that we represented the posterior probability by sampling 50 circles, the maximum entropy occurs when the proposed measurement location is inside of 25 circles (and outside of 25 circles).  As the robot's parameter estimates converge, one can show that the system is simply performing a binary search by asking `yes' or `no' questions, which implies that each measurement results in one bit of information.  However, in the case where the robot employs an SSF model of the light sensor, the question the robot is essentially asking is more detailed: `To what degree does the circle overlap the light sensor's SSF?'  The answer to such a question tends to provide more information, which significantly improves system performance.  One can estimate the information gain achieved by employing the SSF model.  Consider that na\"{i}ve model reveals that estimating the circle's position and radius with a precision of 4mm given the prior information about the circle and the playing field requires 25 bits of information.  The experiment using the SSF model requires on average 16 measurements, which implies that on average each measurement obtained using the SSF model provides about $25/16 = 1.56$ bits of information.  One must keep in mind, however, that this is due to the fact that the SSF model is being used not only to infer the circle parameters from the data, but also to select the measurement locations.

Because the method presented here is based on a very general inferential framework, these methods can easily be applied to other types of sensors and equipment in a wide variety of situations.  If one has designed a robotic machine learning system to rely on likelihood functions, then sensor models can be incorporated in more or less a plug-and-play fashion.  This not only promises to improve the quality of robotic systems forced to rely on lower quality sensors, but it also opens the possibility for calibration on the fly by updating sensor models as data are continually collected.

\section*{Acknowledgements}
This research was supported in part by the University at Albany Faculty Research Awards Program (Knuth) and the University at Albany Benevolent Research Grant Award (Malakar).  We would like to thank Scott Frasso and Rotem Guttman for their assistance in interfacing MatLab to the LEGO Brick.  We would also like to thank Phil Erner and A.J. Mesiti for their preliminary efforts on the robotic arm experiments and sensor characterization.

\bibliographystyle{IEEEtran}
		 \bibliography{references}
\end{document}